# Fine-structure Constant, Anomalous Magnetic Moment, Relativity Factor and the Golden Ratio that Divides the Bohr Radius

R. Heyrovska (1) and S. Narayan (2)

((1) J. Heyrovsky Institute of Physical Chemistry, Prague, Czech Republic and (2) Villa Julie College, Stevenson, MD, USA )

Sommerfeld introduced the fine-structure constant ($\alpha$) into physics, while he was taking into account the relativistic effects in the theory of the hydrogen atom. Ever since, it has puzzled many scientists like Eddington, Dirac, Feynman and others. Here the mysterious fine-structure constant, $\alpha$ ($= \lambda_{C,H}/\lambda_{dB} = 1/137.036 = 2.627/360$) is interpreted based on the finding that it is close to $\phi^2/360$ ($= 2.618/360 = 1/137.508$), where $\lambda_{C,H}$, the Compton wavelength for hydrogen is a distance equivalent to an arc length on the circumference $\lambda_{dB}$ ($= 2\pi a_B$) of a circle with the Bohr radius ($a_B$) and $\phi$ is the Golden ratio, which was recently shown to divide the Bohr radius ($a_B$) into two Golden sections at the point of electrical neutrality. From the data for the electron ($e^-$) and proton ($p^+$) g-factors, it is found that $(360/\phi^2) - \alpha^{-1} = 2/\phi^3 = (g_p - g_e)/(g_p + g_e)$, where $360/\phi^2 = \lambda_{dB}/(\lambda_{C,H} - \lambda_{C,H,i})$, $\lambda_{C,H,i} = \phi 2\pi r_{\mu,H}$ and $r_{\mu,H}$ is the sum of the intrinsic radii of $e^-$ and $p^+$ evaluated from the g-factors and that $\alpha - \phi^2/360 = 0.009/360 = \lambda_{C,H,i}/\lambda_{dB} = (1- \gamma)/\gamma$, the factor for the advance of perihilion in Sommerfeld's theory of the hydrogen atom, where $\gamma$ is the relativity factor.

This world year of physics, WYP 2005, is also concerned with the mystery of the fine-structure constant ($\alpha$) which puzzled Feynman [1]: *"... is it related to pi or perhaps to the base of natural logarithms? Nobody knows. It's one of the greatest damn mysteries of physics: a magic number that comes to us with no understanding by man. You might say the*

*"hand of God" wrote that number, ..."* , see Gilson [2a]. $\alpha$ and its relation to the g-factors are recently rousing considerable interest [2 - 5]. The constancy (or not) of $\alpha$ defined as $2\pi e^2/\kappa hc$, where e, $\kappa$, h and c are respectively, the charge, electric permittivity of free space, Planck's constant and the speed of electromagnetic waves in vacuum, is being debated mostly in terms of these parameters. The present relation between $\alpha$ and the magnetic moment anomalies expressed as a long series of terms [6] is too complicated to comprehend. This work investigates $\alpha$ from a different point of view, on observing that it is close to $\phi^2/360$ (= 1/137.508), where $\phi$ is the Golden ratio [7], also called the Divine ratio (Feynman thought of $\pi$ or e!), which was recently found [8,9] to play an important role in atomic physics.

$\phi$ is a mathematical constant which operates in a wide variety of geometrical workings in the Universe ranging from mollusks in the sea to the spiral galaxies in the cosmos [7]. Geometrically, when the segments a and b of a line of length d are such that (a + b)/a = a/b, the ratio a/b = $(1 + 5^{1/2})/2 = \phi = 1.618033989...$ , and a (= d/$\phi$) and b (= d/$\phi^2$) are called the Golden sections. Nature works through the remarkable and unique property of $\phi$ that the power series, ..., $1/\phi^2$, $1/\phi$, 1, $\phi$, $\phi^2$, $\phi^3$, ..., is at the same time also a Fibonacci series (0, 1, 1, 2, 3, 5, 8, etc.), where any Fibonacci number is related to its neighbors as follows: e.g., 3 = 1 + 2 = 5 – 2. Thus, $\phi = (1/\phi) + 1 = \phi^2 - 1$. Also, note that $\phi^2/\pi = 0.833346 \sim (5/6)$.

In the case of a hydrogen atom, which consists of an electron (e⁻) and a proton (p⁺) (with magnetic moments $\mu_e$ and $\mu_p$ respectively) separated by the Bohr radius $a_B$, it was shown recently [8] that the ground state energy ($E_H$) is given by the difference,

$$E_H = (1/2)(e^2/\kappa a_B) = (1/2)(e^2/\kappa)[(1/a_{B,p}) - (1/a_{B,e})] = eI_H = h\nu_H \qquad (1a)$$

$$1/a_B = (1/a_{B,p}) - (1/a_{B,e}); \quad a_B = a_{B,e} + a_{B,p}; \quad (a_{B,e}/a_{B,p})^2 - (a_{B,e}/a_{B,p}) = 1 \qquad (1b\text{-}d)$$

where $I_H$ is the ionization potential of hydrogen, $I_e = -e/(\kappa a_{B,e})$ and $I_p = e/(\kappa a_{B,p})$ are the component potentials of $(e^-)$ and $(p^+)$ respectively, $\nu_H = c/\lambda_H$ is the frequency of light that ionizes hydrogen and $\lambda_H$ is the corresponding wavelength. From Eqs. (1b,c), one obtains the Golden quadratic equation Eq. (1d) whose positive root is $a_{B,e}/a_{B,p} = \phi$. This shows that $a_B$ is divided at the point of electrical neutrality $P_{el}$ into the two Golden sections, $a_{B,e} = a_B/\phi$ and $a_{B,p} = a_B/\phi^2 = a_{B,e} - a_B/\phi^3$.

The above interpretation in terms of the Golden ratio was extended to other atoms and further, it was shown [8] that for any two atoms of the same kind, the covalent bond distance is the sum of its Golden sections, the larger one being the anionic and the smaller one, the cationic radius of the atom. This enabled assignment of ionic radii for atoms, and many bond lengths were shown to be additive. E.g., the crystal ionic distances of alkali halides could be quantitatively reproduced by adding the respective ionic radii.

Since the de Broglie wavelength, $\lambda_{dB}$ ($= 2\pi a_B$) is equal to the circumference of the circle with the radius $a_B$, it is also the sum of the circumferences of two Golden circles, $\lambda_{dB,e} = 2\pi a_{B,e} = (2\pi a_B)/\phi$ and $\lambda_{dB,p} = 2\pi a_{B,p} = (2\pi a_B)/\phi^2$ with the radii $a_{B,e}$ and $a_{B,p}$. These are also equal to the Golden arc lengths corresponding to the Golden angles, $(2\pi/\phi)$ and $(2\pi/\phi^2)$ radians or $360°/\phi$ ($= 222.492°$) and $360°/\phi^2$ ($= 137.508°$) respectively. Note: $2\pi$ radians ($= 2\pi r/r$) are divided into 360 degrees where both the radian and the degree are dimensionless.)

$\alpha$ ($= 2\pi e^2/\kappa hc = 1/137.036 = 2.627/360$) is the ratio of wavelengths,

$$\alpha = \lambda_{C,H}/\lambda_{dB} = \lambda_{dB}/(\lambda_H/2) \tag{2}$$

where $\lambda_{C,H} = \lambda_{C,e} + \lambda_{C,p}$, the sum of the Compton wavelength shifts for the electron and proton, corresponds to an arc length, subtended by a central angle of $360/137.036 = 2.627°$ ($= \phi^2 + 0.009$), of the circumference $\lambda_{dB}$ of the circle. Since $\alpha = \lambda_{C,H}/\lambda_{dB} = r_{C,H}/a_B$, it is the ratio of $r_{C,H}$ (= $\lambda_{C,H}/2\pi$), the Compton radius, which is related to the magnetic moments and $a_B$, which is associated with the electrical charges of $e^-$ and $p^+$. $r_{C,H}$ is given by,

$$\mu_B + \mu_N = (e/2\mu_{H,red})h_{bar} = (ec/2)r_{C,H} \tag{3a}$$

$$h_{bar} = h/2\pi = cm_e r_{C,e} = cm_p r_{C,p} = c\mu_{H,red}\, r_{C,H} \tag{3b}$$

$$\mu_B/r_{C,e} = \mu_N/r_{C,p} = (\mu_B + \mu_N)/r_{C,H} = ec/2 \tag{3c}$$

where $\mu_B = (ec/2)r_{C,e} = (e/2m_e)h_{bar}$ is the Bohr magneton and $\mu_N = (ec/2)r_{C,p} = (e/2m_p)h_{bar}$ is the nuclear magneton, $r_{C,H} = r_{C,e} + r_{C,p}$, $m_e$ and $m_p$ are the rest masses of the electron and proton, respectively, and $\mu_{H,red} = m_e m_p/(m_e + m_p)$ is the reduced mass of hydrogen. Eq. (3c) shows that the magnetic pole strength is $ec/2$.

The g-factor ($g_e/2$) for the electron is defined as the ratio $\mu_e/\mu_B$, and is given by,

$$\mu_e = (g_e/2)\mu_B = (1 + a_e)\mu_B = \mu_B + \mu_{e,intr} = (ec/2)R_{\mu,e} \tag{4a}$$

$$g_e/2 = R_{\mu,e}/r_{C,e} = (c/h_{bar})m_e R_{\mu,e} = I_{\mu,e}/I \tag{4b}$$

where $\mu_{e,intr} = a_e\mu_B = (ec/2)r_{\mu,e}$ can be considered as the intrinsic magnetic moment of the electron, $r_{\mu,e}$ is the intrinsic radius, $a_e = r_{\mu,e}/r_{C,e}$, is the electron magnetic moment anomaly [6b] and $R_{\mu,e} = r_{C,e} + r_{\mu,e}$. Eq. (4b) shows that $g_e/2$ is the ratio of the moments $I_{\mu,e} = m_e R_{\mu,e}$ (due to the magnetic moment) and $I = h_{bar}/c$.

Similarly, $\mu_N$ and $\mu_p$ are related by $g_p/2$ as shown,

$$\mu_p = (g_p/2)\mu_N = (1 + a_p)\mu_N = \mu_N + \mu_{p,intr} = (ec/2)R_{\mu,p} \tag{5a}$$

$$g_p/2 = R_{\mu,p}/r_{C,p} = (c/h_{bar})m_p R_{\mu,p} = I_{\mu,p}/I \tag{5b}$$

5where $\mu_{p,intr} = a_p\mu_N = (ec/2)r_{\mu,p}$ is the intrinsic magnetic moment of the proton, $r_{\mu,p}$ is the intrinsic radius, $a_p = r_{\mu,p}/r_{C,p}$ is the proton magnetic moment anomaly and $R_{\mu,p} = r_{C,p} + r_{\mu,p}$. Eq. (5b) shows that $g_p/2$ is proportional to the moment, $I_{\mu,p} = m_p R_{\mu,p}$.

On dividing $a_B$ into the sections $a_{\mu,e}$ and $a_{\mu,p}$ at a point, $P_\mu$ [8,9] such that the moments of inertia (due to the magnetic moments), $I_{\mu,e}a_{\mu,e} = I_{\mu,p}a_{\mu,p}$, one obtains,

$$I_{\mu,e}/I_{\mu,p} = m_e R_{\mu,e}/m_p R_{\mu,p} = m_e\mu_e/m_p\mu_p = a_{\mu,p}/a_{\mu,e} = g_e/g_p = 0.3585 \tag{6a}$$

$$(a_{\mu,e} - a_{\mu,p})/a_B = (g_p - g_e)/(g_e + g_p) = 2/\phi^3 = 2(a_{B,e} - a_{B,p})/a_B = 360/\phi^2 - \alpha^{-1} \tag{6b}$$

In Eq. (6b), use of the data from [6c] for the g-factors gives $(g_p - g_e)/(g_e + g_p) = 0.472 = 2/\phi^3$ and therefore, $g_e/g_p = (\phi^3 - 2)/(\phi^3 + 2)$; see also [9].

From Eqs. (2) and (4) - (6), one finds that $360/\phi^2$ stands for the ratio,

$$360/\phi^2 = \lambda_{dB}/(\lambda_{C,H} - \lambda_{C,H,i}) \tag{7}$$

where, $\lambda_{C,H,i} = \phi 2\pi r_{\mu,H}$ and $r_{\mu,H} = r_{\mu,e} + r_{\mu,p}$. The distances, $\lambda_{C,H}$, $(\lambda_{C,H} - \lambda_{C,H,i})$ and $\lambda_{C,H,i}$ correspond to small arc lengths on the circle of circumference $\lambda_{dB}$, subtended by central angles of $2.627^0$, $2.618^0$ ($= \phi^2$) and $0.009^0$ ($= 2.627 - 2.618$) respectively.

On noting that $0.009^0 = 360(1 - \gamma)/\gamma = 0.009(6)^0$, the advance of the perihelion in Sommerfeld's theory [10] of the hydrogen atom, where $\gamma = (1 - \alpha^2)^{1/2} = 0.99997(3)$ is the relativity factor when the angular momentum $p = h_{bar}$, one obtains the following relations,

$$\lambda_{C,H,i}/\lambda_{dB} = \alpha - \phi^2/360 = (1 - \gamma)/\gamma \text{ and } \gamma = \lambda_{dB}/(\lambda_{dB} + \lambda_{C,H,i}) = 0.99997(5) \tag{8a,b}$$

55

Thus the Golden ratio provides a quantitative link between the various known quantities in atomic physics.

Figure 1 shows the role of ϕ in the hydroegn atom and the caption gives all the details.

One of us (R. H.) is thankful to the Ministry of Industry and Trade of the Czech Republic (project No. 1H-PK/42) for the financial support.

**Figure 1.** The Golden ratio, point of electrical neutrality ($P_{el}$) and magnetic center ($P_\mu$) of the hydrogen atom.

AB = $a_B$ (= $5.29465 \times 10^{-11}$m, from the data in [6c]), the Bohr radius, is the distance between the electron (e⁻) with magnetic moments $\mu_e$ at A and proton (p⁺) with magnetic moments $\mu_p$ at B, and O is the center of AB. The (auxiliary) right-angled triangle ABC is constructed for locating the point of electrical neutrality, $P_{el}$, such that BC = BO = $a_B/2$ = CD. This makes, AC = $\sqrt{5}a_B/2$ and AD = $a_B/\phi$, where $\phi$ = 1.618034. $P_{el}$ is then marked on AB so that $AP_{el}$ = AD = $a_B/\phi$ = $a_{B,e}$ and $BP_{el}$ = $a_{B,p}$ = $a_B/\phi^2$.

EA = $AP_{el}$ = $a_B/\phi$ = $0.618a_B$ and BG = $BP_{el}$ = $a_B/\phi^2$ = $0.382a_B$. F is the center of EG (= $2a_B$) and $2FO = 2OP_{el} = FP_{el} = (a_{B,e} - a_{B,p}) = a_B/\phi^3 = 0.236a_B$, The circumference of the circle EHGJE with F as the center = $2\pi a_B = \lambda_{dB}$, the de Broglie wavelength. The Golden angles HFJ = $360/\phi^2$ (= $137.508°$) and $360/\phi$ (= $222.492°$). The Golden arcs HGJ = $(2\pi/\phi^2)a_B$ = $\lambda_{dB,p}$ and HEJ = $(2\pi/\phi)a_B = \lambda_{dB,e}$. Note that a common tangent to these two circles from a point beyond G (not shown in Fig. 1), will be inclined to AB at an angle $\sin^{-1}(a_{B,e} - a_{B,p})/a_B$ = $\sin^{-1}(1/\phi^3)$.

$P_\mu$, the magnetic center, is located such that $(g_e/g_p) = a_{\mu,p}/a_{\mu,e}$, $2P_{el}P_\mu = FP_{el} = a_B/\phi^3$, $AP_\mu = a_{\mu,e} = a_B/2 + (a_B/\phi^3) = (\phi^3 + 2)/2\phi^3 = 0.736a_B$, $BP_\mu = (g_e/g_p)a_{\mu,e} = a_{\mu,p} = a_B/2 - (a_B/\phi^3) = (\phi^3 - 2)/2\phi^3 = 0.264a_B$ and $(a_{\mu,e} - a_{\mu,p})/a_B = 2/\phi^3 = 360/\phi^2 - \alpha^{-1}$. Note: The chord HJ intersects AB at I close to $P_\mu$ and BI = $0.255a_B$.

The sum of the Compton wavelengths of (e⁻) and (p⁺), $\lambda_{C,H}$ (= $\alpha\lambda_{dB}$), ($\lambda_{C,H} - \lambda_{C,H,i}$) [= $(\phi^2/360)\lambda_{dB}$] and $\lambda_{C,H,i}$ {= $\phi 2\pi r_{\mu,H} = (\alpha - \phi^2/360)\lambda_{dB} = [(1 - \gamma)/\gamma]\lambda_{dB}$, where $r_{\mu,H}$ is the sum of the intrinsic radii of (e⁻) and (p⁺) and $\gamma$ is the relativity factor,} correspond to arc lengths, on the circle EHGJE of circumference $\lambda_{dB}$, which subtend at the center F, the angles, $2.627°$, $2.618°$ (= $\phi^2$) and $0.009°$ respectively.

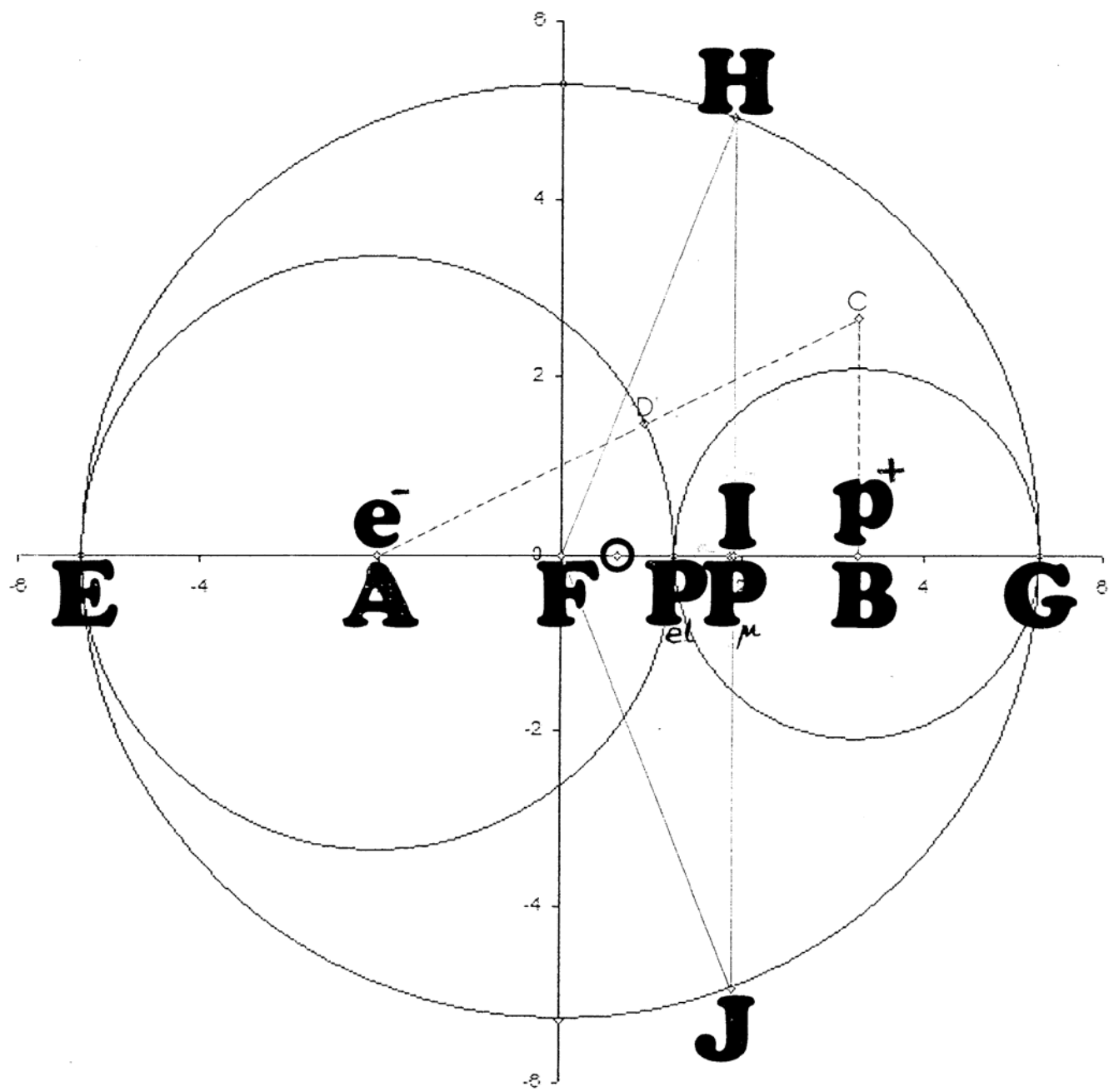